\documentclass[conference]{IEEEtran}
\makeatletter

\def\ps@IEEEtitlepagestyle{%
    \def\@oddfoot{\mycopyrightnotice}%
    \def\@evenfoot{}%
}
\def\mycopyrightnotice{%
    {\footnotesize  978-1-7281-4117-6/19/\$31.00 \copyright2019 IEEE\hfill \\}
     {\footnotesize  DOI 10.1109/DataCom.2019.00029 }
    \gdef\mycopyrightnotice{}
}


\makeatletter
\newcommand*\titleheader[1]{\gdef\@titleheader{#1}}
\AtBeginDocument{%
  \let\st@red@title\@title%
  \def\@title{%
    \bgroup\normalfont\large\centering\@titleheader\par\egroup
    \vskip1.5em\st@red@title}
}
\makeatother
\titleheader{2019 IEEE 5th International Conference on Big Data Intelligence and Computing (DATACOM)}
\title{Distributed-Memory Vertex-Centric Network Embedding for Large-Scale Graphs*\\
\thanks{This work was supported by the NSF CCF Award \#1725585.}}

\IEEEoverridecommandlockouts
\usepackage{algorithm}
\usepackage{algorithmicx}
\usepackage[noend]{algpseudocode}
\usepackage{mathtools}
\usepackage{color}
\usepackage{amsmath}
\usepackage{amsfonts}
\usepackage{booktabs}
\newcommand{\eat}[1]{}
\newcommand{\mbf}[1]{\mathbf #1}
\newcommand{\mc}[1]{\mathcal #1}

\begin{document}

\author{\IEEEauthorblockN{Sara Riazi}
\IEEEauthorblockA{\textit{Department of Computer and Information Science} \\
\textit{University of Oregon}\\
Eugene, OR 97403, USA \\
riazi@uoregon.edu}
\and
\IEEEauthorblockN{Boyana Norris}
\IEEEauthorblockA{\textit{Department of Computer and Information Science} \\
\textit{University of Oregon}\\
Eugene, OR 97403, USA \\
bnorris2@uoregon.edu}
}

\maketitle

\begin{abstract}
Network embedding is an important step in many different computations based on graph data. 
However, existing approaches are limited to small or middle size graphs with fewer than a million edges. 
In practice, web or social network graphs are orders of magnitude larger, thus making most current methods impractical for very large graphs.
To address this problem, we introduce a new distributed-memory parallel network embedding method based on Apache Spark and GraphX. We demonstrate the scalability of our method as well as its ability to generate meaningful embeddings for vertex classification and link prediction on both real-world and synthetic graphs. 
%To the best of our knowledge, we are the first to train network embeddings for graphs with billions of edges.

\end{abstract}

%\keywords{Network embedding, Apache Spark}
\IEEEpeerreviewmaketitle

\section{Introduction}

Network embedding is an important step in solving many graph problems including link prediction, vertex classification, and clustering. Network embedding aims to learn a low dimensional vector representation for vertices of a graph. However, existing approaches do not scale to very large graphs with billions of vertices and edges. One solution is to use distributed-memory systems and out-of-core computation.

Among distributed-memory systems, frameworks such as the Apache Spark-based GraphX~\cite{graphx} are of particular interest to us because they offer a map-reduce-based approach to expressing parallel algorithms for graph computations.

In order to take advantage of such distributed graph processing frameworks, we need to design new map-reduce~\cite{mapreduce} network embedding algorithms. In general, following the previous work for learning general network embedding~\cite{deepwalk,line,node2vec}, we use the structural properties of a network to train an embedding. A common assumption underlying existing methods and our new algorithm is that we expect that the embedding of a vertex is more similar to the embeddings of its neighbors rather than to the embedding of a random vertex outside of its neighborhood.  We enforce this objective with approximate maximum likelihood training of the embedding in which the partition function is approximated using negative samples. This training requires lookup access to the embedding of vertices in a neighborhood, as well as vertices that lie outside of the neighborhood. However, lookup access in map-reduce frameworks is prohibitively expensive, which necessitates careful consideration in developing map-reduce based network embedding algorithms. In this paper, we introduce such an algorithm, and experimentally show that we can train network embeddings for very large graphs. We evaluate the new algorithm's accuracy and parallel scalability on a set of real-world networks.

Our \textbf{key contributions} include the following.
\begin{itemize}
    \item A discussion of the limitations of GraphX for implementing existing network embedding algorithms.
    \item A new map-reduce-friendly message propagation model for learning vertex-centric network embeddings, which propagates the gradients instead of the embedding.  
    \item The use of random graphs to construct negative sampling, which is necessary for approximate maximum likelihood training. 
    \item A new GraphX based vertex-centric network embedding (VCNE) algorithm based on gradient propagation and random graphs that performs well on a range of real-world problems and synthetic graphs and can be applied to large problems that cannot be handled by current embedding approaches.
\end{itemize}

\section{Parallel Graph Frameworks}
Applying traditional graph algorithms to extremely large graphs requires distributed processing as well as out-of-core computation. Therefore, several parallel graph frameworks such as GraphX~\cite{graphx} and Giraph~\cite{giraphpp} have been developed on top of data-parallel systems, such as Apache Spark and Hadoop, respectively. As a result, they provide graph processing APIs using distributed data processing models such as map-reduce~\cite{mapreduce}. In map-reduce, data is converted to key-value pairs and then partitioned onto nodes. A map-reduce system consists of a set of workers that are coordinated by a master process. The master process assigns partitions to workers, and then workers apply a user-defined map function to the key-value pairs, resulting in intermediate key-value pairs stored on the local disks of workers. Apache Spark defines the map-reduce model in term of operation over distributed collection objects called resilient distributed datasets (RDDs). RDDs~\cite{rdd} are immutable collections of objects that are partitioned across different Spark nodes in the network. 

An RDD is transformed into another RDD using transformation instructions, such as {\em map} and {\em filter}.  
Transformations in Spark are lazy, which means that Spark does not apply transformations immediately. Instead, it constructs a directed acyclic graph (DAG) of data parts and transformations followed by final steps as actions. Then it executes the formed DAG by sending it as several tasks to Spark nodes. The actions in Spark reduce RDDs to values. For example, {\em count} computes the number of records in RDDs, so it needs all the transformation to be applied first, and then it returns the result. 
%GraphX defines a graph structure using vertex and edge RDDs and provide high-level API such as Pregel programming model using transformations and actions over these RDDs.

Because Spark is a data-parallel computation system, GraphX implements graph operations based on the data-parallel operations available in Spark, such as join, map, and reduce. GraphX represents graphs using two RDDs, one for vertices and another for edges. 

However, handling graphs in a data-parallel computation system is more complex than map-reduce operations since the vertices should be processed in the context of their neighbors. To address that, GraphX introduces edge triplets, which join the structure of vertices and edge RDDs. Each triplet carries an edge attribute and the attributes of vertices incident to that edge. Therefore, by grouping the triplets on the id of the source or destination vertex, one can access the value of all the neighbors of each vertex. Moreover, since the triplets are distributed, if the neighbors of a vertex are located on different machines, then Spark workers have to communicate with each other to construct the result. Therefore, different strategies for distributing graphs over partitions result in significant differences in communication and storage overheads. GraphX supports both edge-cut and vertex-cut graph partitioning strategies. 

GraphX also provides a vertex-centric programming model for developing distributed graph algorithms. Vertex-centric programming models such as Pregel~\cite{pregel} are widely used for reimplementing sequential algorithms in graph-parallel frameworks such as Apache Giraph or GraphX. In a vertex-centric programming model, we develop an algorithm from a vertex point of view, which in general includes three different steps: gathering messages from its neighboring vertices, updating its state, and generating messages for its neighbors. The graph-parallel framework iteratively executes these steps in one super-steps until no more messages are produced by any vertex. GraphX implements all this functionality using map-reduce operations over edge triplets. 

\section{Vertex-Centric Network Embedding}
The goal of vertex-centric network embedding is to learn a low-dimensional vector for each vertex in the graph such that the vector representation carries the structural properties of the graph.
Formally, for graph $G(\mc{V}, \mc{E})$ of vertex set $\mc{V}$ and edge set $\mc{E}$, we want to learn a $d$-dimensional vector representation $\mbf{u}_i$ for each $i\in \mc{V}$ such that $d \ll |\mc{V}|$.

Existing approaches to learning vector representations~\cite{deepwalk,line,node2vec,graphsage,gat} aim to encode the neighborhood of a vertex (its structural properties) into a low-dimensional space. 
Other properties of vertices,  such as attributes, labels, and relations can also be incorporated into the vector representation of the vertex~\cite{embprop,lin2015learning,tadw,tri-party}.

In general, a graph embedding approach is vertex-centric-friendly if the embedding of each vertex is a function of only the embeddings of its neighbors. 
For example, LINE-1st~\cite{line} computes the embedding using first-order proximity by optimizing the following objective function:
\begin{align}
    \max_\mbf{u}\sum_{(i,j)\in E} w_{ij} \sigma(\mbf{u}_i^T \mbf{u}_j), \label{rel:line1}
\end{align}
in which $\mbf{u}_i$ and $\mbf{u}_j$ are vector representations of vertex $i$ and $j$, respectively, $\sigma$ is a sigmoid function, and $w_{ij}$ is the edge weight.
We can rewrite Eq.~\ref{rel:line1} as 
\begin{align}
    \max_\mbf{u} \sum_i \sum_{j\in N(i)}  w_{ij} \log \sigma(\mbf{u}_i^T \mbf{u}_j), \label{rel:line1-r}
\end{align}
where $N(i)$ is the set of neighbors of vertex $i$. 

More powerful representation learning methods, such as LINE second-order proximity defines the conditional probability of vertex $v_j$ as the context of the vertex $v_i$: 
\begin{align}
    p(v_j|v_i) = \frac{\exp(\mbf{u}_j^T\mbf{u}_i)}{\sum_{j=1}^{|V|}\exp(\mbf{u}_j^T\mbf{u}_i)},
\end{align}
where $|V|$ is the number of vertices in the graph, and minimizes the KL-divergence between empirical distribution $\frac{w_{ij}}{w_i}$ and $p(v_j|v_i)$, where $w_i = \sum_j w_{ij}$. This is resulted in the following optimization problem:
\begin{align}
    \sum_{(i,j)\in E} \frac{w_{ij}}{w_i}\log p(v_j|v_i). \label{rel:o2}
\end{align}
Similarly this equation can also be written in a vertex-centric manner by factorizing over vertices:
\begin{align}
    \sum_i \sum_{j\in N(i)} \frac{w_{ij}}{w_i} \log p(v_j|v_i).
\end{align}
The main problem arises in the computing of the normalization part of $p(v_j|v_i)$:
\begin{align}
  \log p(v_j|v_i) = \mbf{u}_j^T\mbf{u}_i - \log \sum_{j=1}^{|V|}\exp(\mbf{u}_j^T\mbf{u}_i)  
\end{align}

The term $\log \sum_{j=1}^{|V|}\exp(\mbf{u}_j^T\mbf{u}_i)$ is intractable to compute, but can be estimated using negative samples. However, in order to have a vertex-centric approximation, we replace it with $\sum_{j=1}^{d_i} \mbf{u}_j^T\mbf{u}_i$, which is a lower bound of $\log \sum_{j=1}^{|V|}\exp(\mbf{u}_j^T\mbf{u}_i)$ using Jensen's inequality.
Therefore our final objective function becomes:
\begin{align}
\max_u \sum_i \frac{1}{w_i} \sum_{j\in N(i)} w_{ij} \mathbf{u}_i^T\mathbf{u}_j +  \sum_{j\notin N(i)}^{d_i}  - \mathbf{u}_i^T\mathbf{u}_j.~\label{rel:line2}
\end{align}
 
The above objective enforces the similarity if  the embeddings of neighbors \emph{and} the dissimilarity of embeddings of random vertices selected among non-neighbor nodes (negative samples), contrasting them to learn the embedding of each vertex.

Negative samples ensure that the objective function does not find a trivial solution (e.g., the embeddings of all vertices become the same). Negative sampling simply forces the embeddings of non-neighbor nodes to be different. 

In a vertex-centric paradigm, we are required to decompose the algorithm such that each vertex is responsible for its part of the objective function evaluation, providing all the necessary information, e.g., the current state of its neighbors. In other words, we look at the computation from a vertex point of view.
We can simply view network embedding of Eq.~\ref{rel:line2} in a vertex-centric paradigm: ``As a vertex, I want my embedding to be similar to my neighbors' embeddings, while it differs from the embeddings of other non-neighbor vertices.''
% A vertex-centric network embedding requires the objective function to decompose as partial objectives computable at individual vertices, but unfortunately the objective of Eq.~\ref{rel:line2} does not decompose over vertices.

%  We can generalize this view point to different network embedding algorithms. A vertex-centric network embedding requires the objective function to decompose as partial objectives, each respecting a vertex:

In a vertex-centric setting for optimizing based on Eq.~\ref{rel:line2}, each vertex needs to access the embeddings of vertices that are not directly connected to it (negative sampling). Parallel graph frameworks do not provide efficient lookup of random vertices that are distributed among different machines. Moreover, each compute node does not have a lookup dictionary that can be used to locate and ship the attributes of required vertices, but there are routing tables for vertices based on the edges that are connecting them, so accessing the neighboring vertices is efficient (compared to random lookup access).

To benefit from this efficiency, we define a random graph, in which each vertex $i$ is connected to $d_i$ randomly selected vertices in the graph with a negative weight, where $d_i$ is degree of vertex $i$. We construct a new graph as the union of the current graph and the random graph. In the new augmented graph, each vertex has access to the embedding of $d_i$ randomly chosen vertices. Therefore, we can rewrite Eq.\ref{rel:line2} with our augmented graph, decomposed over the vertices:
\begin{align}
O_i = \max_u \sum_{j\in A(i)} w_{ij} \mathbf{u}_i^T\mathbf{u}_j, \label{rel:oi}
\end{align}
where $w_{ij}$ is negative one for negative neighbors and the weight of the connecting edge for the actual neighbors, and $A(i)$ is the set of neighbors of vertex $i$ in the augmented graph. 
% We can derive Eq.~\ref{rel:oi} from Eq.~\ref{rel:line2} by using the symmetry in the sigmoid function: $\sigma(-x) = -\sigma(x)$ and absorbing $k$ in the weights.

The objective function of Eq.~\ref{rel:oi} decomposes over vertices in the augmented graph, so it can be computed in a vertex-centric approach unlike the negative sampling-based approach in the original graph, whose objective function is not decomposable. 

After each step, we also normalize the embedding so every embedding has a norm of one. This make sure that the magnitude of embedding remains bounded, so that the contribution of vertices in the objective function is similar.

\subsection{Vertex-Centric algorithm}
A data-parallel vertex-centric graph algorithm typically involves three steps: sending messages among neighbors (sendMessage), reducing all the messages to a single vertex to one message (mergeMessage), and executing a vertex related function given the final reduced message and the current state of the vertex (vertexProgram). The sendMessage is emulated by mapping each triplet (joint data structure of an edge and the vertices that are incident with the edges) into a set of messages. Each message has a key that determines the vertex id of its destination. The data-parallel engine takes every pair of messages that are targeted for the same vertex and reduces them into one message that can be merged with another message for the same vertex. Finally, at most one message is left for the target vertex. The vertexProgram takes that message and the state of the target vertex and produces a new state for the target vertex. Here, for example, the state is the embedding of a vertex. The message generation requires  map-reduce operation over triplets, while state update requires join operation among the old vertex RDD and the message RDD. We must keep the size of the intermediate structures such as messages constant with respect  to the number of neighbors, otherwise for graphs with power-law degree distributions, the message size may become prohibitively large.

% In a vertex centric paradigm, we decompose the algorithm such that each vertex is responsible for its part providing the the necessary information, for example, the currant state of its neighbors. In other words, we look at the computation from a vertex point of view.

% We can simply view network embedding in a vertex centric paradigm: "As a vertex, I want my embedding to be similar to my neighbors while it differs other non-neighbor vertices". We can generalize this view point to different network embedding algorithms. A vertex-centric network embedding requires the objective function to decompose as partial objectives, each respecting a vertex: % of Eq.~\ref{rel:line1} over vertices:
% \begin{align}
% O_i = \sum_{j\in N(i)} \sigma(\mbf u_i^T \mbf u_j) - \frac{1}{k} \sum^k_{j \in \text{NegN}(i)}  \sigma(\mbf u_i^T \mbf u_j), \label{rel:oi}
% \end{align}

% \begin{align}
%     \mbf m_{i\leftarrow j} = \mbf u_j \\
%     Reduce(m_{i\leftarrow j}, m_{i\leftarrow k}) = \{m_{i\leftarrow j},m_{i\leftarrow k}\} \\
%     O_i = \sum_j \sigma(e_w * \mbf u_i^T m_{i\leftarrow j}) \\
%     \mbf u_i = \mbf u_i + \eta \nabla_i O_i 
% \end{align}
% where $NegN(i)$ is a set of k vertices that are not neighbor with vertex $i$.
In order to compute the partial objective $O_i$ on each compute node, a naive implementation sends the embedding of each neighbor to vertex $i$ as sendMessage, keeps the union of embeddings as the reduceMessage, and optimizes $O_i$  in the vertexProgram. However, in a map-reduce framework, combining the embedding vectors can result in prohibitively large collections since there is no bound on the degree of the vertices. 

This large collection are constructed in the mergeMessage steps since the vertexProgram is executed only when all the messages have been passed. 
% We can avoid the construction of the large collections by introducing 

We use a simple trick to avoid the construction of these large collections by propagating the gradient instead of the embeddings. However, we first have to make sure that the total gradient of Eq.~\ref{rel:oi} can be computed by the vertex programs. 
% Knowing that $-\sigma(x)=\sigma(-x)$, we can rewrite eq.~\ref{rel:oi} with one sum over the augmented neighborhood $A(i) = N(i) \bigcup 
% \text{NegN}(i)$ in which the edge weight of negative samples are $-1$:
% % \begin{align}
% % O_i = \sum_{j\in A(i)} w_{ij} \sigma(e_w*\mbf u_i^T \mbf u_j). \label{rel:oi}
% % \end{align}

The gradient of $O_i$ can be written as 
\begin{align}
\nabla O_i = \sum_{j\in A(i)} \nabla O_{i\gets j}, \label{rel:gradient}
\end{align}
where 
\begin{align}
    \nabla O_{i\gets j} = w_{ij} * \mbf u_j \label{rel:edge_grad}
\end{align}
\\
Finally we can update the embedding using gradient ascent:
\begin{align}
    \mbf u_i = \mbf u_i + \eta \nabla O_i \label{rel:gradient_update}
\end{align}

Using edge triplets, each vertex in the augmented neighborhood $A(i)$ has access to data structures needed for computing $\nabla O_{i\gets j}$. Therefore, defining $\nabla O_{i\gets j}$ as a sendMessage function and sum as the mergeMessage operation, the final reduced message for vertex $i$ is Eq.~\ref{rel:gradient}. Finally, vertexProgram executes the gradient update. 
In this vertex-centric design, the size of the data structures remains bounded and no large collection would be constructed in the intermediate steps. Therefore, we can optimize Eq.~\ref{rel:oi} for very large graphs with large vertex degrees. Algorithm~\ref{alg1} shows the definition of these functions.
% GraphX runs the message-passing using triplets that augment structure containing the attributes of an edge and vertices that are incident with the edge, which enables the neighboring vertices $N(i)$ to compute their share of gradient for optimizing $O_i$. 

\begin{algorithm}[t]

\caption{Vertex-Centric Network Embedding}\label{alg1}
\begin{algorithmic}
\State //$e_{ji}$ : edge from $j$ to $i$.
\State //d: embedding dimension
\State //msg: (m: $|\mathbb{R}|^d$)
\State //vertex attributes: ($u$: $|\mathbb{R}|^d$)
\State //$m_{i \rightarrow j}$: means the message from $i$ for $j$ 
\Procedure{sendMessasge}{$e_{ij}$, $u_i$, $u_j$}
\State $m_{j \rightarrow i}: \nabla O_{i\gets j}$ //Eq.~\ref{rel:edge_grad}
\EndProcedure

\Procedure{MergeMessages}{$m_{i \rightarrow j}$, $m_{k \rightarrow j}$}
\State $m_{i \rightarrow j}$ + $m_{k \rightarrow j}$ //Eq.~\ref{rel:gradient} 
\EndProcedure

\Procedure{vertexProgram}{$u$, $m$}
\State $u \gets u + \eta m$ // Eq.\ref{rel:gradient_update}
\EndProcedure
\end{algorithmic}

\label{alg:vcne}
\end{algorithm}

\section{Experiments}
We compare our network embedding algorithm, VCNE, with LINE~\cite{line}, Node2vec~\cite{node2vec} and PyTorch-BigGraph~\cite{pbg} on medium-size datasets to show the capability of VCNE to learn meaningful representation. Then, we apply VCNE to very large graphs for the task of link prediction. Table~\ref{tab:graphs} reports the characteristics of the graphs used in our experiments.

\begin{table}[h!tb]
\caption{The number of vertices and edges of the real-world graphs in our test suite.} %The RMAT examples show the sizes of RMAT-ER. Other RMAT types have comparable sizes and are not listed due to space limitations.
\label{tab:graphs}
\centering
\begin{tabular}{l r  r}
\toprule
Name&  Num. of Vertices & \quad Num. of Edges  \\
 \midrule
% RMAT & 339,201,984 & 4,252,445,904 \\
Friendster & 68,349,466	 & 2,586,147,869  \\
Twitter-MPI & 52,579,682 & 1,963,263,821  \\
Twitter & 41,637,597 & 1,453,833,084   \\
LiveJournal & 5,193,874 &  48,682,718  \\
Reddit & 232,965 & 11,606,919\\ 
PPI & 56,944 & 793,632 \\
% Cora & 23,145& 90,568 \\
\bottomrule
\end{tabular}
\end{table}

\subsection{Vertex Classification}
The vertex classification goal is to classify each vertex into different groups, which includes both multi-class and multi-label classification.

We use two datasets of protein-protein interaction (PPI) and Reddit posts. In PPI, the goal is to assign a set of activated protein functions to each vertex, which are represented using positional gene sets, motif gene sets, and immunological
signatures~\cite{graphsage}. The total number of possible protein functions is 121, and the vertex feature set size is 50. 

Reddit is an online discussion forum in which people publish posts and comment on others' posts. In the Reddit graph, the vertices are the posts and two vertices are adjacent if a user comments on the posts corresponding to the vertices~\cite{graphsage}. The node features include the average word embedding of the title, all comments of the post and the score of the post as well as the number of comments on the posts. The total number of features is 602, and the goal is to assign each vertex to one of 41 communities.  
For both PPI and Reddit, we used the same set of train/val/test as provided by \cite{graphsage}. Table~\ref{tab:graphs} shows the characteristics of these two graphs.

We first generate vertex embeddings using LINE, Node2Vec, Pytorch-BigGraph and VCNE. Next, we concatenate the vertex embedding to the vertex features and use it as input to a logistic regression classifier to predict labels. As a baseline, we also train logistic regression using only the vertex features.
Although more complex classifiers such as multi-layer perceptron would be possible and may result in higher accuracy, we use simple logistic regression to better isolate the impact of vertex embedding.

We used an embedding dimension of 100 for all algorithms.

\begin{table}[thb]
    \centering
    \caption{$F_1$ score of vertex classification tasks using different embedding algorithms.}
    \label{tab:my_label}
    \begin{tabular}{l r  r}
    \toprule
        & PPI & Reddit \\
        \hline
    Vertex features & 43.3 &  51.2 \\
    % LINE (1st)  &  &  \\
    % LINE (2nd) & & \\
    LINE  & 53.08  &  63.9 \\
    Node2vec & 49.8 & 65.4 \\
    % MILE & 48.9 & 65.9\\
    PyTorch-BigGraph & 52.70  & 66.3 \\
    VCNE & \textbf{53.28} & \textbf{66.7} \\

    \bottomrule
    \end{tabular}
    \label{tab:vertex_classifer}
\end{table}

\begin{figure}[h!tb]
    \centering
    \includegraphics[width=0.9\columnwidth]{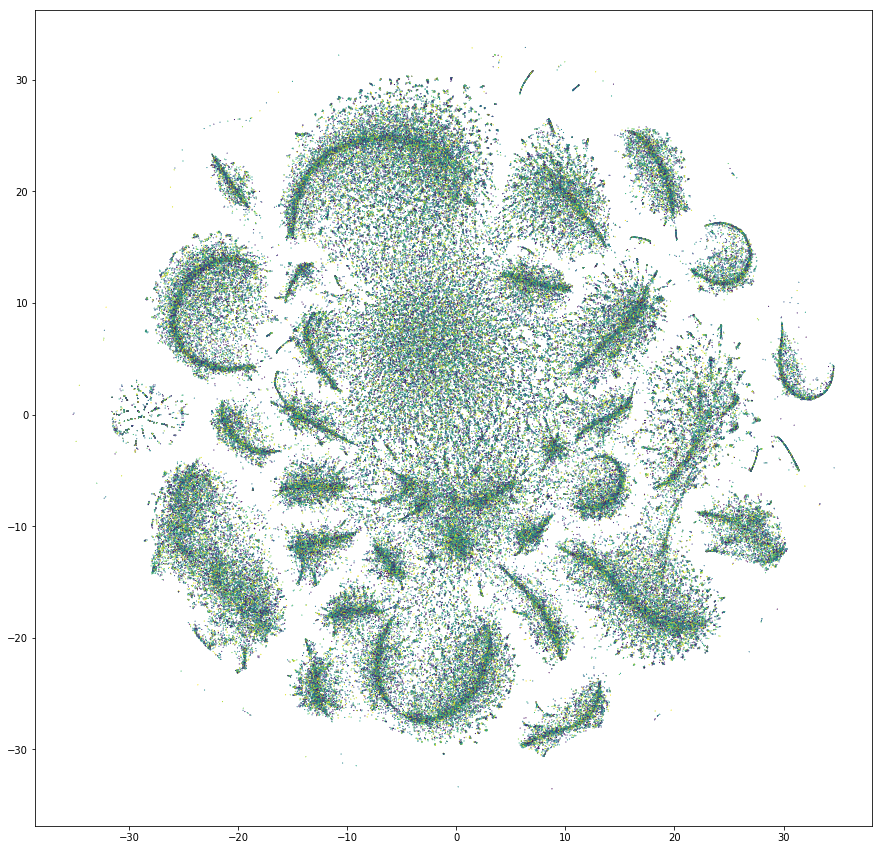}
    \caption{The embedding of the Reddit graph generated by VCNE.}
    \label{fig:embedding}
\end{figure}

Table~\ref{tab:vertex_classifer} shows the performance VCNE, LINE, Node2Vec, and raw vertex features in terms of their $F_1$ score.
For all embedding algorithm, using the embedding in addition to vertex features helps, so we can conclude that the embedding is meaningful and encodes structural properties of the graph. For  both Reddit and PPI, VCNE is more accurate than all the baselines. We also show the learned embedding by VCNE using t-SNE~\cite{t-sne} in Figure~\ref{fig:embedding}. VCNE can capture clear clusters in the graph. 

\subsection{Link Prediction}
Link prediction is an important graph analytic problem, in which we wish to predict the potential edges in the network. This problem is of particular interest for social network friend suggestion or predicting future evolution of graphs.

We constructed a synthetic link prediction dataset, for which we dropped one percent of the current edges of the graph and kept the dropped edges as the test set combined with another set of vertex pairs as the true negative. The size of our negative set is equal to the size of the dropped set making sure that we have a balanced test set. We generate the training and validation sets using the same approach. The remaining edges of the graph constitute the core graph, on which the network embedding algorithms have been trained. We emphasize that the training algorithms \emph{have not seen} the dropped edges.
We first compare LINE, PyTorch-BigGraph and VCNE on the LiveJournal graph.

\begin{table}[h!tb]
    \centering
     \caption{Link Prediction for LiveJournal}
    \begin{tabular}{l r r r}
    \toprule
         & Precision& Recall & \(F_1\) \\
         \hline
         Jaccard & 99.9 &  82.6 &90.4\\ 
         LINE &  90.8 & 84.9 & 87.8\\
         Pytorch-BigGraph & 92.0 & 80.7& 86.0 \\ 
         VCNE &  93.3 & 88.1 & 90.6 \\
         \bottomrule
    \end{tabular}
 
    \label{tab:link-lj}
\end{table}

% \begin{table}[h!tb]
%     \centering
%      \caption{The performance of link prediction using VCNE.}
%     \begin{tabular}{l r r r}
%     \toprule
%          & Precision & Recall & \quad $F_1$ \quad\\
%          \hline
%          Friendster& 69.3 & 98.0 & 81.2  \\
%          Twitter MPI &67.8 &  94.7 &79.0 \\
%          Twitter & 72.4 &  93.9 &  81.7  \\
%          LiverJournal & 67.3 &  81.7 &  73.8\\
%          \bottomrule
%     \end{tabular}
 
%     \label{tab:link}
% \end{table}

% \begin{table}[h!tb]
%     \centering
%      \caption{Link prediction comparison with 2-layer mlp}
%     \begin{tabular}{l r r}
%     \toprule
%          & VCNE& Pytorch-BigGraph \\
%          \hline
%          Friendster&  &  \\
%          Twitter MPI &  \\
%          Twitter &  85.2 & 92.6   \\
%          LiverJournal & & \\
%          \bottomrule
%     \end{tabular}
 
%     \label{tab:link}
% \end{table}

\begin{table}[h!tb]
    \centering
     \caption{The performance of link prediction using VCNE}
    \begin{tabular}{l|c|c|c}
    \toprule
         & Precision & Recall & $F_1$ \\
         \hline
         Friendster& 84.8& 93.5& 88.9   \\
         Twitter MPI &87.5 & 84.4& 85.9 \\
         Twitter & 80.7 &90.0 & 85.1   \\
         \bottomrule
    \end{tabular}
 
    \label{tab:link}
\end{table}

We also use Jacard index to predict an edge: $J(u,v) = \frac{N(u) \cap N(v)}{N(u)\cup N(v)}$, where $N(u)$ is the set of neighbors of vertex $u$.
Computing the Jacard index requires constructing triplets whose vertex attributes are sets of neighbor IDs, and for very large social networks, this results in prohibitively large messages given the power-law degree distribution of social networks. 
Nevertheless, we could compute the Jacard index for LiveJournal graphs, but not for the other larger graphs. The cut threshold for deciding the existence of an edge is selected based on the validation data. For LiveJournal, using the Jacard index results in 99.2\% precision, 71.1\% recall, and $F_1$ score of 83.1\%. 
For the link prediction using embeddings, we train a 2-layer multi-layer perceptron with 500 hundred hidden units using the training pairs. We pick the best model based on the performance on the validation set, and report the model performance on the test set.

Table~\ref{tab:link-lj} the performance of link prediction for LiveJournal graph. The Jaccard index has the highest precision, while VCNE has the best performance in overall $F_1$ score. 

% \subsection{Fixed random graph vs }
Next, we apply VNCE to the very large graphs that cannot be handled by other approaches and report the results in Table~\ref{tab:link}. For all graphs, the $F_1$ score is above 85\%.
% We report the performance of link prediction using the embedding generated by VCNE in 

\subsection{Scalability}
To measure the parallel scalability of VCNE over Apache Spark, we run VCNE for Friendster, Twitter MPI, Twitter, and LiveJournal with different numbers of Spark workers: 10, 20, 30, and 40. Each worker has access to 20 cores and 75 GB of memory (for a total number of cores ranging between 200 and 800 and memory ranging from 750 GB to 3 TB). The University of Oregon Talapas cluster, on which we performed the experiments, consists of dual Intel Xeon E5-2690v4 nodes connected with an EDR InfiniBand network.

\begin{figure}[h!tb]
    \centering
    \includegraphics[width=1.0\columnwidth]{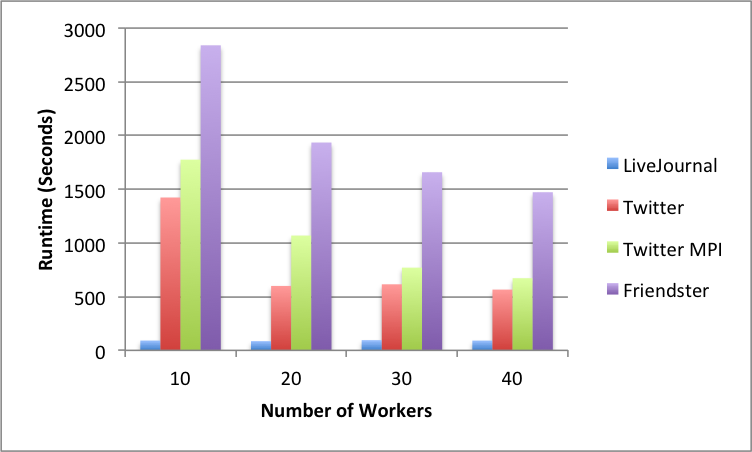}
    \caption{Average runtime for one training step of VCE with 10 to 40 Spark workers. }
    \label{fig:scale}
\end{figure}

Figure~\ref{fig:scale} reports the average runtime for one learning iteration, which includes generating the random graph, combining the random graph with the original graph, and updating the embedding using Algorithm~\ref{alg1}.
We observe that the overhead of using data-parallel systems such as Apache Spark for processing mid-size graphs such LiveJournal is considerable, but increasing the number of workers significantly helps the processing of larger graphs such as Twitter-MPI and Friendster.

We also study the effect of the dimension of embedding and the number of negative samples on the running time of VCNE.
\begin{figure}
    \centering
    \includegraphics[width=\columnwidth]{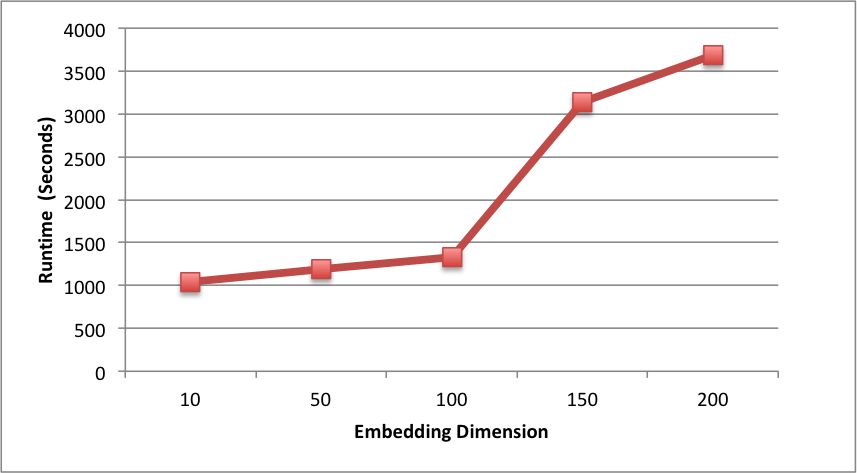}
    \caption[The effect of embedding dimension on the running time.]{The effect of embedding dimension on the running time for the Livejournal graph.}
    \label{fig:lj_emb}
\end{figure}
These two factors directly affect the performance of the underlying map-reduce implementation. As we increase the dimension of embedding the local memory required for distributed map-reduce operations increases, thus imposing more overhead on the system. We measure the running time of 10 iterations of training VNCE for the Livejournal graph. We used 10 workers with 20 cores and 80 GB of memory each. Figure~\ref{fig:lj_emb} reports the results, which shows the running time of VCNE with respect to the dimension of embedding.

\begin{figure}
    \centering
    \includegraphics[width=\columnwidth]{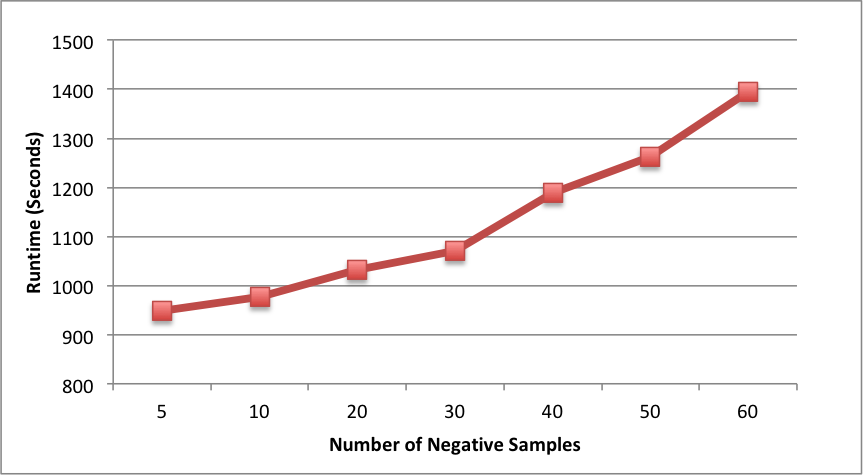}
    \caption[The effect of the number of negative samples on the running time.]{The effect of the number of negative samples on the running time for the Livejournal graph.}
    \label{fig:lj_neg}
\end{figure}

We also study the effect of negative sampling on the running time of VCNE on the Livejournal graph with different numbers of negative samples. Negative samples increase the size of augmented graph, thus increasing the number of messages and the running time (see Figure~\ref{fig:lj_neg}).

\subsection{Implementation Details}
Working with iterative algorithm over very large graphs may result in replicating large collections such as EdgeRDDs in local memory. It is very important to unpersist the collection from memory in order to avoid exceeding the available memory capacity. For example, in the pipeline operations such graph construction followed by groupEdge, Apache Spark materializes the first graph and we lose the pointer to it as it is followed by map operation. It is necessary to observe the storage memory profile provided by Apache Spark as a part of its Web UI to make sure that no large collections are left behind in an iteration. 

We observe that unpersisting the RDDs may not force freeing the memory, and some RDDs may continue to reside in the memory waiting for the garbage collector. This behavior becomes critical for iterative algorithms: increasing the memory usage and activating out-of-core processing, when it is not truly necessary. Therefore, to enforce evacuating the memory, we serialize the working RDDs and close the Spark session at the end of each iteration. This trick is not necessary for mid-size graphs, however, for the consistency we apply it all of the reported experiments.  

Moreover, operations such as aggregateMessage, which are used for message passing over graphs requires significant amount of data shuffling for shipping vertex attributes (embeddings) among workers. This results in a large amount of out-of-core data, which is stored in local storage accessible to the workers, and limits the size of vertex attributes given a fixed number of workers. 

% It is also worth mentioning that RDD element lookup is very expensive. Also, collecting the embeddings of large graphs into the Spark driver is not feasible; hence, we had to implement evaluation metrics such as $F_1$ score using similar vertex-centric distributed-memory parallel approach as the embedding computation.

% \subsubsection{Evaluation}

% \begin{figure}
%     \centering
%     \includegraphics[width=0.5\textwidth]{cora_link.png}
%     \caption{Link prediction performance on Cora citation graph.}
%     \label{fig:my_label}
% \end{figure}

% \begin{figure}
%     \centering
%     \includegraphics[width=0.7\columnwidth]{twitter_link.png}
%     \caption{Link prediction performance on Twitter graph.}
%     \label{fig:my_label}
% \end{figure}

% \begin{figure}
%     \centering
%     \includegraphics[width=0.7\columnwidth]{lj_link.png}
%     \caption{Link prediction performance on LiveJournal graph.}
%     \label{fig:my_label}
% \end{figure}

\section{Related Work}

Many previous works study network embedding~\cite{line,sdne,node2vec,graphsage,embprop}. However, none of these approaches can handle very large graphs. Some previous work has algorithmic restrictions for scaling to very large graphs: for example, SDNE~\cite{sdne} learns low-dimensional embedding using autoencoders, and DeepWalk~\cite{deepwalk} uses hierarchical softmax to parameterize the probability distribution of a vertex given its neighbors.
LINE and Node2Vec do not suffer from algorithmic restrictions, but reimplementing their algorithms for very large graphs is not trivial.\footnote{The authors of Node2vec also provide an Apache Spark implementation, but the implementation is prohibitively slow for very large graphs.} 
The recently introduced PyTorch-BigGraph~\cite{pbg}, however, can be executed for large graphs. PyTorch-BigGraph partitions the vertices into groups, and then partitions edges into buckets based on the groups of vertices that each edge connect. PyTorch-BigGraph then runs traditional network embedding algorithm for buckets that do not share vertex groups in parallel. It selects negative samples from the same vertex groups of the same bucket.

Embeddings can be trained for task-specific purposes by propagating the supervision signal from task loss, e.g., in vertex classification.
Many semi-supervised learning algorithms can be reduced to vertex classification~\cite{semi-graph}. The graph captures the similarity among the points, so every vertex represents one data point, which is either labeled or unlabeled. This problem is also known as collective classification~\cite{collective}. Several methods have been proposed for collective classification, e.g., iterative classification~\cite{collective}. Label Propagation~\cite{lp} is another well-known algorithm for vertex classification.

Graph Convolutional Networks~\cite{gcn}, Graph Attention Networks~\cite{gat}, and GraphSage~\cite{graphsage} are trained by using this supervised signal. These algorithms are not designed to learn embeddings for general purposes and are not scalable to very large graphs.

\section{Conclusions}
We introduced a new distributed-memory parallel vertex-centric algorithm for learning network embeddings of very large graphs using GraphX and Apache Spark. Our algorithm, VCNE, can easily scale to handle very large graphs (billions of vertices and edges or larger) by increasing the number of Apache Spark workers. We also show the VCNE can learn meaningful representations as demonstrated by the performance of two use cases, classification and link prediction. %% Future work?

% \section{Acknowledgments}
% This research was supported by NSF CCF Award \#1725585. 
%
% The next two lines define the bibliography style to be used, and the bibliography file.
\bibliographystyle{ieeetran}
\bibliography{graph}

% 
% If your work has an appendix, this is the place to put it.

\end{document}